\documentstyle[preprint,pra,epsfig,aps,wick]{revtex}

\newcommand\beq{\begin{equation}}
\newcommand\eeq{\end{equation}}
\newcommand\bea{\begin{eqnarray}}
\newcommand\eea{\end{eqnarray}}

\newcommand{\ket}[1]{| #1 \rangle}
\newcommand{\bra}[1]{\langle #1 |}

\newcommand{\ba}{\begin{array}}
\newcommand{\ea}{\end{array}}
\newtheorem{theo}{Theorem}

\begin{document}
\draft

\title{Classical simulation of noninteracting-fermion quantum circuits}

\author{Barbara~M.~Terhal and David~P.~DiVincenzo}

\address
{IBM T.J. Watson Research Center, P.O. Box 218, Yorktown Heights, NY 10598 USA}

\maketitle

\begin{abstract}
We show that a class of quantum computations that was recently shown
to be efficiently simulatable on a classical computer by Valiant
\cite{valiant:simulation} corresponds to a physical model of
noninteracting fermions in one dimension. We give an alternative proof
of his result using the language of fermions and extend the result to
noninteracting fermions with arbitrary pairwise interactions, where
gates can be conditioned on outcomes of complete von Neumann
measurements in the computational basis on other fermionic modes in the
circuit. This last result is in remarkable contrast with the case of
noninteracting bosons where universal quantum computation can be
achieved by allowing gates to be conditioned on classical bits
\cite{klm}.
\end{abstract}


\section{Introduction}
To understand the power of a quantum computer, it is worthwhile to
explore under what restrictions that power is weakened so as to make
the computation efficiently simulatable with the use of a classical
device. A nontrivial example is the Knill-Gottesman class of quantum
computations \cite{gott:heis}, which can be shown to be efficiently
simulatable on a classical computer. In the quantum circuit we only
allow 1-qubit Hadamard transformations, 1-qubit $\pi/2$ phase shifts,
1-qubit Pauli-rotations and 2-qubit CNOT gates and furthermore the
final measurements are projections in the two eigenspaces 
of any sequence of Pauli-matrix observables.

This question about restricted classes of quantum computation is also
related to the question of universality of a quantum computation. What
set of gates, or in more physical terms, what physical system can be
used to implement universal quantum computation? Surprises have been
found in this direction, for example it was shown that the $2$-qubit
exchange interaction is sufficient for universal computation and, for
example, that universal computation can be achieved with a network of
phase shifters, beamsplitters and photon counters, i.e. noninteracting
bosons, where logical gates can be conditioned on previous measurement
outcomes \cite{klm}.

In Ref. \cite{valiant:simulation} a new class of quantum computations
is introduced that are shown to be efficiently simulatable on a
classical device.  The class includes a special set of unitary 2-qubit gates
on nearest-neighbor qubits.
In this paper we will analyze this class of gates and show that it maps onto a system of noninteracting fermions (i.e., associated with Hamiltonian interactions that are quadratic in fermion creation and annihilation operators) in one dimension.  The equivalence will enable us to give a straightforward derivation of the classical simulation, as well as extend the class of quantum
computations to include (1) noninteracting fermions without nearest
neighbor restrictions and (2) gates that are applied conditionally on
measurement outcomes. In particular the second result, when compared
with the universal computation by linear optics in Ref. \cite{klm},
shows a fundamental difference between bosons and fermions. One
possible cause for the difference is that the bosonic modes, unlike
fermionic modes, can contain more than one particle, a feature that is
employed in the universality construction in Ref. \cite{klm}.

The class of gates that Valiant shows to be classically simulatable
is larger than the class that we start from in Theorem \ref{theoval} (see below); his class includes non-unitary gates and also some special set of 2-qubit 
(possibly non-unitary) gates on the first 2 qubits. E. Knill has now shown 
\cite{knill:flo} that this entire class is (indeed) weaker than full quantum computation. Furthermore, Knill shows 
that the extensions that we treat in this paper to non-nearest neighbor interactions and conditionally applied gates are in fact included in Valiant's class 
of gates, albeit in a nonconstructive manner. 

In Section \ref{nonint} we establish the mapping from Valiant's gate
set to a system of fermions. In Section \ref{preserve} we show how the
classical simulation comes about when we restrict ourselves to
quadratic interactions that preserve the fermion number and in Section
\ref{gensim} we handle the general case of noninteracting
fermions. Finally, in Section \ref{ccgates} we show how classically
conditioned gate operations can likewise by simulated with our methods.

\section{Noninteracting Fermions}
\label{nonint}

Let us first state the main theorem of Ref. \cite{valiant:simulation};
we will give a slightly restricted version of the theorem that does
not include the extra freedom of gate choice on the first two qubits
nor the possibility to do non-unitary gates:

\begin{theo}[Valiant \cite{valiant:simulation}]
Let $M$ be the unitary transformation representing a quantum circuit on 
$n$ qubits that consists of $2$-qubit gates $U$ on qubits $x_i$ and $x_{i+1}$, $i=0,\ldots,n-1$, where $e^{i\phi}U$ is of the form 
\beq
e^{i\phi}U=\left(
\ba{llll}
U_{11}^1 & 0 & 0 & U_{12}^1 \\
0 & U_{11}^2 & U_{12}^2 & 0 \\
0 & U_{21}^2 & U_{22}^2 & 0 \\
U_{21}^1 & 0 & 0 & U_{22}^1
\ea
\right),
\eeq 
where $U^1$ and $U^2$ are arbitrary elements of $SU(2)$ and $\phi$ is
an arbitrary phase. There exist polynomial-time classical algorithms
that evaluate (1) $|\bra{y} M\ket{x}|^2$ for arbitrary bitstrings $x$
and $y$, (2) ${\rm Tr} \bra{y^*} M \ket{x}\bra{x}M^{\dagger}\ket{y^*}$
where $y^*$ corresponds to an assignment of an arbitrary $k$-bit subset
for any $k$, and (3) sample, given an arbitrary input string
$\ket{x}$, the probability distribution over outcomes $y^*$ of a
measurement (in the computational basis) on an arbitrary $k$-bit
subset of the qubits.
\label{theoval}
\end{theo}

Note that (3), which corresponds to the final simulation of a quantum
computation, follows from (2) in a fairly straightforward manner (see
Ref. \cite{valiant:simulation}) whereas (2) could be strictly stronger
than (1).

A first observation about the class of allowed gates is that they preserve 
the parity of an input bitstring $\ket{x}$, which is expressed by 
the fact that the $\{\ket{00},\ket{11}\}$ sector is decoupled from the 
$\{\ket{01},\ket{10}\}$ sector. Note that the overall phase factor $e^{i\phi}$ 
is irrelevant in the computation.

To make contact with physical models, we write a gate $U$ acting on
nearest-neighbor qubits $i$ and $i+1$ as $e^{iH}$.  This
Hamiltonian $H$ can be written as a sum of three types of interactions:
\beq
H_1=\alpha_1 Z_i \otimes {\bf I}_{i+1}+\beta_1 {\bf I}_{i} \otimes Z_{i+1},
\label{h1}
\eeq
\beq
H_2=\alpha_2 X_i \otimes X_{i+1}+\beta_2 Y_i \otimes Y_{i+1},
\label{h2}
\eeq
and 
\beq
H_3=\alpha_3 X_i \otimes Y_{i+1}+\beta_3 Y_i \otimes X_{i+1}.
\label{h3}
\eeq
where $\alpha_j,\beta_j$ are real, and $X,Y,Z$ are the three Pauli matrices:
\bea
X=\left(
\ba{lr}
0 & 1 \\
1 & 0  
\ea
\right),\; & 
Y=\left(
\ba{lr}
0 & -i \\
i & 0  
\ea
\right),\; &
Z=\left(
\ba{lr}
1 & 0 \\
0 & -1  
\ea
\right).
\eea

At this point we note that the gate set in Theorem \ref{theoval} seems
extremely close to a universal set of gates. It has been proved
\cite{baconetal} that universal quantum computation can be achieved by
employing {\em only} the $XY$-interaction, i.e. $H \propto X\otimes X+Y\otimes Y$, {\em if these gates can be applied on any pair of
qubits}. Since this form of interaction is certainly allowed in Theorem
\ref{theoval}, we conclude that the nearest-neighbor constraint is
crucial in the construction.  Another observation is that adding
arbitrary 1-qubit gates to this gate-set would also result in
universality; it has been proved that universal quantum computation
can be obtained with a circuit with arbitrary 1-qubit gates and only
nearest-neighbor (1 dimensional) $XY$-interactions \cite{imametal}.

Let us now consider the mapping onto a system of fermions. We can
identify the $n$-bit computational basis states $\ket{x}$ with a state
of $n$ fermionic modes, each of which can be occupied, corresponding
to $1$, or unoccupied, corresponding to $0$. We have a set of
operators, creation operators $a_i^{\dagger}$ and (hermitian
conjugate) annihilation operators $a_i$ associated with each mode $i$
which obey the anticommutation rules
\bea
\{a_i,a_j\}\equiv a_i a_j+a_j a_i=0,\;\;&\;\; \{a_i^{\dagger},a_j^{\dagger}\}=0,\;\; &\;\; \{a_i,a_j^{\dagger}\}=\delta_{ij}I.
\label{ac}
\eea
The annihilation and creation operators act on computational basis states 
in the following manner, consistent with the anticommutation relations:
\beq
a_i \ket{x_0,\ldots x_i,\ldots,x_{n-1}}=\delta_{x_i,1}e^{i \pi \oplus_{m=0}^{i-1}x_i} \ket{x_0,\ldots,\bar{x_i},\ldots,x_{n-1}},
\eeq
and 
\beq
a_i^{\dagger} \ket{x_0,\ldots x_i,\ldots,x_{n-1}}=\delta_{x_i,0}e^{i \pi \oplus_{m=0}^{i-1}x_i} \ket{x_0,\ldots,\bar{x_i},\ldots,x_{n-1}}.
\eeq
Given these definitions, we can transform the Pauli operators in
Eqs. (\ref{h1}-\ref{h3}) to creation and annihilation operators of
fermions by a Jordan-Wigner transformation
\cite{jwtrafo,pliberg}. This is done by first defining the operators
$\sigma_i^\pm={1\over 2}(X_i\pm iY_i)$ that relate to the annihilation
and creation operators as
\bea
\sigma_j^+=e^{i \pi \sum_{m=0}^{j-1}a_m^{\dagger}a_m}a_j^{\dagger},\;\;&\;\;
\sigma_j^-=e^{i \pi \sum_{m=0}^{j-1}a_m^{\dagger}a_m}a_j.
\label{jw}
\eea
With these rules, the three types of interactions $H_1,H_2$ and $H_3$ can be 
rewritten as (we omit terms that are proportional to $I$ since they will only
add irrelevant phase factors to the quantum state of the computer):
\beq
H_1=2\alpha_1  a_i^{\dagger}a_i+2\beta_1 a_{i+1}^{\dagger}a_{i+1},
\eeq
\beq
H_2=\alpha_2(a_i^{\dagger}-a_i)(a_{i+1}^{\dagger}+a_{i+1}) 
-\beta_2(a_i^{\dagger}+a_i)(a_{i+1}^{\dagger}-a_{i+1}), 
\eeq
and 
\beq
H_3=-i \alpha_3 (a_i^{\dagger}-a_i)(a_{i+1}^{\dagger}-a_{i+1})
-i \beta_3(a_i^{\dagger}+a_i)(a_{i+1}^{\dagger}+a_{i+1}).
\eeq
Thus we see that the total Hamiltonian $H=H_1+H_2+H_3$ is a sum of
nearest-neighbor fermionic interactions that are quadratic in the
fermion creation and annihilation operators, i.e., we can obtain any
real linear combination of the Hermitian operators $a_i^{\dagger}
a_i$, $a_{i+1}^{\dagger}a_{i+1}$,
$a_i^{\dagger}a_{i+1}-a_ia_{i+1}^{\dagger}$,
$i(a_i^{\dagger}a_{i+1}+a_ia_{i+1}^{\dagger})$,
$a_{i}^{\dagger}a_{i+1}^{\dagger}-a_{i}a_{i+1}$, and
$i(a_{i}^{\dagger}a_{i+1}^{\dagger}+a_{i}a_{i+1})$.  Fermionic systems
that evolve according to such a quadratic Hamiltonian are referred to
as `noninteracting', because a canonical transformation (change of
basis) exists that brings the Hamiltonian into a standard form
involving a sum of terms each of which acts only on a single mode.

We note that if the initial gate set in terms of Pauli matrices did
not have the nearest-neighbor restriction, then the corresponding
fermion interaction would not have been quadratic: this is due to the
nonlocal `sign' part in the Jordan-Wigner transformation,
Eq. (\ref{jw}).  It has been found that these nonlocal signs are not a
problem when one considers the question of simulating the dynamics of
fermionic systems on a quantum computer: fermions dynamics can be
simulated efficiently on a quantum computer, see
Refs. \cite{al},\cite{kitaev:fermi} and \cite{ortizetal}.
Furthermore, it has been shown in Ref. \cite{kitaev:fermi} that
universal quantum computation can be obtained by fermionic
interactions that include Hamiltonians that are {\em quartic} in the
annihilation and creation operators.  Terms with an odd number of
fermion operators are unphysical (they could transform an isolated fermion
into an isolated boson), but they have some interesting mathematical
features, see the discussion, Section \ref{disc}.

\section{Preserving the number of fermions}
\label{preserve}

Before we discuss how a fermionic circuit involving $H_1$, $H_2$ and
$H_3$ can be simulated classically, we show how this simulation is
done in the more restricted case when the gates preserve the number of
fermions. Thus we consider a circuit on $n$ fermionic modes where each
elementary gate $U$ corresponds to an interaction between modes $i$
and $j$, and can be written as $U=\exp(iH_g)$ where the gate
Hamiltonian is written generally as
\begin{equation}
H_g=b_{ii} a_i^{\dagger}a_i+b_{jj} a_j^{\dagger}a_j+ b_{ij}
a_i^{\dagger}a_j+b_{ij}^* a_j^{\dagger}a_i.\label{aexpr}
\end{equation}
Note that the coefficients $b_{\alpha\beta}$ form a $2\times 2$
Hermitian matrix; we will consider these coefficients to be part of an
$n \times n$ matrix {\bf b}, which is only non-zero for matrix
elements involving modes $i$ and $j$.  Here and later in Section
\ref{gensim} we impose no restriction that $i$ and $j$ be
nearest-neighbor modes, unlike the case that Valiant introduced.  We
will abbreviate the vacuum state $\ket{00\ldots 0}$ as $\ket{{\bf
0}}$.  Let $U=U_{{\rm poly}(n)}\ldots U_2U_1$ be a sequence of
$2$-qubit gates representing the quantum circuit.  We consider
\beq
U a_i^{\dagger} \ket{{\bf 0}}=U a_i^{\dagger} U^{\dagger} U \ket{{\bf 0}}=U a_i^{\dagger} U^{\dagger} \ket{{\bf 0}},
\label{pres}
\eeq
since $U\ket{{\bf 0}}=\ket{{\bf 0}}$ due to fermion number preservation.
$U$ acts by conjugation as 
\beq
U a_i^{\dagger} U^{\dagger}=\sum_m V_{im} a_m^{\dagger}.
\label{conju}
\eeq 
When $U$ corresponds to a gate operation as in Eq. (\ref{aexpr}), the
matrix $V$ is given by $V=\exp(i\bf{b})$.  This result is proved by
making a canonical transformation that diagonalizes {\bf b}.  By group
composition, the matrix $V$ for the entire circuit $U$ is given by
matrix multiplication of the $V$s for each gate.  This evaluation of
$V$ is polynomially efficient in $n$ if the circuit contains ${\rm
poly}(n)$ gates (In fact, we could replace the individual $2$-qubit 
gates by an arbitrary quadratic fermion-number preserving Hamiltonian and 
the matrix $V$ of the total circuit could still be evaluated efficiently).

We will first show how to evaluate efficiently the matrix element
$\bra{y}U \ket{x}$ where $\ket{x}$ and $\ket{y}$ are arbitrary input
and output bitstrings.  Since $U$ preserves the number of fermions,
$\bra{y}U\ket{x}=0$ if $x$ and $y$ have different Hamming weight. Let
$U$ act on a state $\ket{x}$ with $k$ fermions in certain positions:
\beq
U \ket{x}=U a_{i_1}^{\dagger} a_{i_2}^{\dagger}\ldots a_{i_k}^{\dagger} \ket{{\bf 0}},
\eeq
where $i_1 < i_2 < \ldots < i_k$ by convention. Using Eqs. (\ref{pres},\ref{conju}), we can write
\beq
U\ket{x}=\sum_{p_1,\ldots,p_k} V_{i_1,p_1}V_{i_2,p_2} \ldots V_{i_k,p_k} a_{p_1}^{\dagger} a_{p_2}^{\dagger}\ldots a_{p_k}^{\dagger} \ket{{\bf 0}},
\eeq 
The output state equals $\bra{y}=\bra{{\bf 0}} a_{l_k} \ldots
a_{l_1}$, where $l_1 < l_2 < \ldots < l_k$.  Using the anticommutation
rules Eq. (\ref{ac}), we see that contributions to the inner product
$\bra{y} U \ket{x}$ only arise when $p_1\ldots p_k$ is some
permutation $\pi$ of the indices $l_1 \ldots l_k$. Furthermore, we get
an overall sign for every such term corresponding to the number of
interchanges of creation operators that we have to perform in order to
rewrite the state $a^{\dagger}_{\pi(l_1)} \ldots
a^{\dagger}_{\pi(l_k)}\ket{{\bf 0}}$ as $a^{\dagger}_{l_1} \ldots
a^{\dagger}_{l_k} \ket{{\bf 0}}$. After this reordering no more sign
changes will take place, since $\bra{{\bf 0}}a_{l_k} \ldots
a_{l_1}a^{\dagger}_{l_1} \ldots a^{\dagger}_{l_k} \ket{{\bf
0}}=\bra{{\bf 0}}a_{l_k}a^{\dagger}_{l_k} \ldots
a_{l_1}a^{\dagger}_{l_1}\ket{{\bf 0}}=1$. Thus
\beq
\bra{y}U\ket{x}=\sum_{\pi} \mbox{ sign}(\pi) 
V_{i_1,\pi(l_1)} V_{i_2,\pi(l_2)}\ldots V_{i_k,\pi(l_k)}.
\eeq
If $\tilde{V}$ is defined as the matrix $V$ where we have selected
rows $i_1,\ldots,i_k$ and columns $l_1,\ldots,l_k$, then we see that
$\bra{y}U\ket{x}=\det(\tilde V)$. The determinant of a $k \times k$
matrix, $k\leq n$, is computable in polynomial time in $n$.

\subsection{Simulating Measurements} 
\label{simmeas}

Next we consider how to simulate classically the outcomes of
measurements on arbitrary subsets of qubits at the end of the
computation. We will show how to calculate the probability that a
certain subset of qubits is in a particular state $y^*$ (item (2) in
Theorem \ref{theoval}). With those probabilities in hand, one can
sample the probability distribution as given by quantum mechanics
(item (3) of Theorem \ref{theoval}).

The Hermitian operator $a_i^{\dagger}a_i$ counts the number of
fermions in mode $i$.  Thus its expectation value with respect to a
density matrix $\rho$, ${\rm Tr}\,a_i^{\dagger}a_i \rho$, is the
probability that mode $i$ is in state $\ket{1}$.  Similarly, the
expectation value of $a_i a_i^{\dagger}=I-a_i^{\dagger}a_i$ is the
probability that mode $i$ is in state $\ket{0}$.  So in order to
evaluate the probability that, given an input state $\ket{x}$, a
certain subset of $k$ modes is in state $\ket{y^*}$ we calculate
\beq
p(y^*|x)={\rm Tr}\,a_{j_1}a_{j_1}^{\dagger} \ldots a_{j_k}^{\dagger}a_{j_k} U \ket{x}\bra{x} U^{\dagger}=\bra{x}\, U^{\dagger}a_{j_1}a_{j_1}^{\dagger} \ldots a_{j_k}^{\dagger}a_{j_k}U \,\ket{x},
\label{pyx1}
\eeq
with $j_1 \neq j_2 \neq \ldots \neq j_k$ and we use $a_i^{\dagger}a_i$
or $a_i a_i^{\dagger}$ when $y_i^*=1$ or $0$ respectively.
(Eq. (\ref{pyx1}) illustrates a case where $y_{j_1}^*=0$ and
$y_{j_k}^*=1$.)  Again we write $\ket{x}=a_{p_1}^{\dagger}\ldots
a_{p_l}^{\dagger} \ket{{\bf 0}}$. We have to evaluate the expression
\bea
\lefteqn{p(y^*|x)=\sum_{m_1,n_1,\ldots,m_k,n_k} V_{j_1,m_1}^{\dagger} V_{n_1,j_1}\ldots 
 V_{j_k,m_k}^{\dagger} V_{n_k,j_k}} \nonumber \\
& &  \bra{{\bf 0}}a_{p_l}\ldots a_{p_1} (a_{n_1}a_{m_1}^{\dagger}\ldots a_{m_k}^{\dagger}a_{n_k}) a_{p_1}^{\dagger}\ldots a_{p_l}^{\dagger} \ket{{\bf 0}}.
\label{pyx}
\eea
We will invoke Wick's theorem \cite{wick} (described in quantum
many-body or quantum field theory textbooks such as
Ref. \cite{mattuck}) to rewrite this formula\footnote{We actually use
Wick's theorem for ordinary operator products rather than for
time-ordered operator products, which is the case analyzed in most
standard treatments.  The version of the theorem we are using is
excellently set out in T. D. Crawford and H. F. Schaefer III, ``An
Introduction to Coupled Cluster Theory for Computational Chemists'',
Reviews in Computational Chemistry {\bf 14}, 33-136 (2000); see also
http://zopyros.ccqc.uga.edu/lec\_top/cc/html/node13.html.}.  Wick's
theorem states that we can rewrite a string of annihilation and
creation operators $A_1 \ldots A_{n}$ as
\beq
A_1 \ldots A_n=\ :A_1 \ldots A_n:+\sum_{k=1}^{\lfloor n/2\rfloor}C_k,
\label{prewick}
\eeq
with
\bea
& &C_1=\ :\wick{1}{<*A_1>*A_2} \ldots A_n:+:\wick{1}{<*A_1 A_2 >*A_3} \ldots A_n:+\ldots \nonumber \\
& &C_2=\ :\wick{1}{<*A_1 >*A_2}\wick{1}{<*A_3 >*A_4} A_5 \ldots A_n:+:\wick{12}{<1A_1 <2A_2 >1A_3 >2A_4}A_5 \ldots A_n:+\ldots  \nonumber \\
& & \mbox{etc.}
\label{wick}
\eea
Here $:A_1 \ldots A_n:$ denotes the so called normal ordered form of
the sequence of operators $A_1 \ldots A_n$.  $:A_1 \ldots A_n:$ is
equal to the reordered sequence of operators $A_{\pi(1)} \ldots
A_{\pi(n)}$ where all the creation operators are moved to the left
(but not reordered amongst each other), and the quantity is negated
when the number of interchanges of creation and annihilation
operators to achieve this form is odd.  The object
$\wick{1}{<*A_{i}>*A_j}$ is called a contraction and is defined as
\beq
\wick{1}{<*A_i >*A_j}=A_i A_j-:A_i A_j:.
\eeq
The terms $C_k$ in Eq. (\ref{prewick}) are each a sum over every
possible choice of $k$ contractions in the normal ordered
product.

 From the anticommutation rules for creation and annihilation operators, 
it follows that
\beq
\wick{1}{<*a^{\dagger}_{i}>*a_j}=\wick{1}{<*a^{\dagger}_{i}>*a_j^{\dagger}}=\wick{1}{<*a_{i}>*a_j}=0,\;\;\wick{1}{<*a_{i}>*a_j^{\dagger}}=\delta_{ij}.
\label{contr}
\eeq
The normal-ordered form is extremely convenient when evaluating an
object such as $\bra{{\bf 0}} A_1 \ldots A_n \ket{{\bf 0}}$ since the
vacuum expectation value of any normal-ordered sequence of operators
vanishes $\bra{{\bf 0}} :A_{i_1}\ldots A_{i_k}:\ket{{\bf
0}}=0$. Therefore, when we evaluate the vacuum expectation value of
Eq. (\ref{wick}), the only terms that remain come from $C_{\lfloor
n/2\rfloor}$ ($n$ even), in which every operator $A_i$ is
contracted (or matched) with another operator $A_j$.  The last step is
to bring the fully contracted terms to a form in which contracted
operators are adjacent, that is, we have
\bea
\lefteqn{\bra{{\bf 0}} :\wick{123}{<1A_1 <2A_2 >1A_3 <3A_4 >2A_5 \ldots >3A_i}\ldots \wick{1}{<*A_{n-1} >*A_n}: \ket{{\bf 0}}=} \nonumber \\
& & {\rm sign}(\pi)\bra{{\bf 0}} :\wick{1}{<*A_1 >*A_3}\wick{1}{<*A_2 >*A_5}\wick{1}{<*A_4 >*A_i}\ldots \wick{1}{<*A_{n-1} >*A_n}: \ket{{\bf 0}}=\mbox{sign}(\pi) \wick{1}{<*A_1 >*A_3}\wick{1}{<*A_2 >*A_5}\wick{1}{<*A_4 >*A_i}\ldots \wick{1}{<*A_{n-1} >*A_n},\label{pyx2}
\eea
where ${\rm sign}(\pi)$ is $-1$ ($1$) when the number of crossings of
the contraction lines is odd (even).  Evidently, what emerges is the
Pfaffian Pf($M$) of $M(i,j)$, an $n \times n$ antisymmetric matrix
(i.e. $M(i,j)=-M(j,i)$).  The Pfaffian Pf($M$) is 0 when $n$ if odd,
and for even $n$ it is defined as \footnote{For an discussion of
Pfaffians via an exposition of Pfaff's 1815 paper introducing the
construction, see T. Muir, {\em The Theory of Determinants in the
Historical Order of Development} (Vol. 1, Dover, 1950), pp. 396-401.}
\beq
{\rm Pf}(M)=\sum_{\pi} {\rm sign}(\pi) M_{\pi(1),\pi(2)} 
\ldots M_{\pi(n-1),\pi(n)},\label{pyx3}
\eeq
where the sum over $\pi$ is restricted to permutations on the indices
$1, 2, \ldots n$ such that
$\pi(2k-1) < \pi(2k)$ and $\pi(1) < \pi(3) < \pi(5)
\ldots$.  Eqs. (\ref{pyx},\ref{pyx2},\ref{pyx3}) tell us that
\beq 
p(y^*|x)={\rm Pf}(M), 
\eeq 
where $M$ can be constructed from Eq. (\ref{pyx}) and the contraction
identities, Eq. (\ref{contr}), in the following manner.  The matrix
elements $M(i,j)$ for $1\leq i < j\leq 2(k+l)$ are obtained from Table
I: The indices $i,j=1, \ldots ,2(k+l)$ are assigned to the ordered
sequence of creation and annihilation operators in Eq. (\ref{pyx}). To
determine $M(i,j)$ ($i < j$) we find what type of operator the indices
$i$ and $j$ correspond to and then read off the matrix element
$M(i,j)$ from the table.  We use unitarity of $V$ and the contraction
rules to determine each entry of the table.  The Xs in the table
indicate that these entries do not occur.

The Pfaffian of an $n \times n$ antisymmetric matrix $M$ can be
computed in poly($n$) time, since ${\rm Pf}(M)^2=\det M$.  The
simulation procedure that was formulated in
Ref. \cite{valiant:simulation} very similarly relies on the evaluation
of a Pfaffian. At the moment it is not clear to us how the representation of the quantum circuit in Ref.  \cite{valiant:simulation} using matchgates 
corresponds to the fermionic representation developed here.

\section{General noninteracting fermions}
\label{gensim}

We are now ready to consider the classical simulation of a quantum
circuit consisting of gates that are built from general quadratic
fermionic interactions. These interactions only preserve the parity of
the photon number. In order to deal with these general interactions,
we transform the set of fermion annihilation and creation operators to
a new set of Hermitian operators (associated with so called Majorana
fermions \cite{kitaev:fermi,kitaev:maj}):
\beq
c_{2i}=a_i+a_i^{\dagger},\;\;c_{2i+1}=-i(a_i-a_i^{\dagger}),
\eeq
where $i=0,\ldots,n-1$. The anticommutation relation for this new set of operators is 
\beq
\{c_k,c_l\}=2\delta_{kl}I.
\eeq
Note that operators $c_{2i}$ and $c_{2i+1}$ are in some sense the fermionic version of conjugate variables $p$ and $q$ that are obtained from linearly combining bosonic annihilation and creation operators.
It is clear that the Hamiltonians $H_1,H_2$ and $H_3$ of
Eqs. (\ref{h1}-\ref{h3}) will be quadratic in these new operators. Let
$U$ be a sequence of $2$-qubit gates each composed of interactions
that are quadratic in the operators $c_i$, i.e. each of the gates corresponds to a Hamiltonian $H$
\beq
H={i\over 4} \sum_{k \neq l} \alpha_{kl} c_k c_l.\label{standard}
\eeq

We again have omitted any term proportional to $I$.  Hermiticity of
$H$ requires only that ${\rm Im}(\alpha_{kl})={\rm Im}(\alpha_{lk})$.
It is conventional to choose the {\boldmath$\alpha$} matrix to be real
and antisymmetric, so that $i${\boldmath$\alpha$} is a Hermitian
$n\times n$ matrix.  For the interactions we have introduced, the
{\boldmath$\alpha$} matrix will only be nonzero in a $4\times 4$
subblock, but this restriction is not necessary for the following
procedure to work.
Similar to the number conserving case, a sequence of gates $U=U_{{\rm poly(n)}}\ldots U_2 U_1$ acts by conjugation as
\beq 
U c_i U^{\dagger}=\sum_j R_{ij}c_j,\label{important}
\eeq 
where $R \in SO(2n)$.  We will establish this important result by
explicitly computing the matrix $R$ for a single gate $U=e^{iH}$.  The result is not so well known
as for the number conserving case (although it has been mentioned in
\cite{kitaev:fermi}), so we will give some of the details of the
derivation.  We follow the notation of \cite{kitaev:maj}.  First, the
Hamiltonian of Eq. (\ref{standard}), with {\boldmath$\alpha$} chosen to
be a real antisymmetric matrix, can be brought into canonical form
\begin{equation}
H={i\over 2}\sum_{j=0}^{n-1}\epsilon_jb'_jb''_j.
\end{equation}
$b'$ and $b''$ are given by the real orthogonal
transformation\footnote{See R. A. Horn and C. R. Johnson, {\em Matrix
Analysis} (Cambridge, 1985), p. 82, Theorem 2.3.4.}
\begin{equation}
\left(\begin{array}{l}b'_0\\b''_0\\ \vdots\\b'_{n-1}\\b''_{n-1}\end{array}\right)=W
\left(\begin{array}{l}c_0\\c_1\\ \vdots\\c_{2n-2}\\c_{2n-1}\end{array}\right).
\end{equation}
The $2n\times 2n$ orthogonal matrix $W$ diagonalizes {\boldmath$\alpha$} into
$2\times 2$ blocks:
\begin{equation}
W{\mbox{\boldmath$\alpha$}}W^T=
\left(\begin{array}{ccccc}0&\epsilon_0&\ &\ &\ \\-\epsilon_0&0&\ &\ &\ \\
\ &\ &\ddots&\ &\ \\ \ &\ &\ &0&\epsilon_{n-1}\\ \ &\ &\ &-\epsilon_{n-1}&0
\end{array}\right).
\end{equation}
The $b'$s and $b''$s have the same anticommutation relations as the original
Majorana fermion operators.  Note that $\pm\epsilon_j$ are the eigenvalues
of the matrix $i${\boldmath$\alpha$}. We now write Eq. (\ref{important}) using the canonical transformation:
\begin{equation}
Uc_iU^{\dagger}=\sum_j\exp(-{1\over 2}\sum_m\epsilon_mb'_mb''_m)
(W_{2j,i}b'_j+W_{2j+1,i}b''_j)\exp({1\over 2}\sum_m\epsilon_mb'_mb''_m).
\end{equation}
Because the Hamiltonian in this canonical form is a sum of commuting terms,
the exponentials here can be factorized; the factors with $m\neq j$ commute 
through and disappear, and we obtain
\begin{equation}
Uc_iU^{\dagger}=\sum_j\exp(-{1\over 2}\epsilon_jb'_jb''_j)
(W_{2j,i}b'_j+W_{2j+1,i}b''_j)\exp({1\over 2}\epsilon_jb'_jb''_j).\label{inter}
\end{equation}
The remaining exponential factors can be expanded and simplified:
\begin{eqnarray}
\exp({1\over 2}\epsilon_jb'_jb''_j)&=&
\sum_{k=0}^\infty{(\epsilon_j/2)^k\over k!}
(b'_jb''_j)^k=\sum_{k=0}^\infty{(\epsilon_j/2)^{2k}\over 2k!}(-1)^k+
\sum_{k=0}^\infty{(\epsilon_j/2)^{2k+1}\over (2k+1)!}(-1)^kb'_jb''_j\nonumber\\
&=&\cos(\epsilon_j/2)+b'_jb''_j\sin(\epsilon_j/2).
\end{eqnarray}
Plugging this form into Eq. (\ref{inter}) and simplifying, we obtain
Eq. (\ref{important}), where $R$ is given by $R=W^TMW$, with the
matrix $M$ having the $2\times 2$ block form
\begin{equation}
M=\left(\begin{array}{rrrrr}\cos\epsilon_0&-\sin\epsilon_0&\ &\ &\
\\ \sin\epsilon_0&\cos\epsilon_0&\ &\ &\ \\ \ &\ &\ddots&\ &\ \\ \ &\
&\ &\cos\epsilon_{n-1}&-\sin\epsilon_{n-1}\\ \ &\ &\
&\sin\epsilon_{n-1}&\cos\epsilon_{n-1}\end{array}\right).
\end{equation}
As before, given a quantum circuit with ${\rm poly}(n)$ two-mode
noninteracting fermion gates, that is, involving four Majorana
fermions per gate, we can construct the total $2n \times 2n$ matrix $R$ in
polynomial time by straightforward matrix multiplication of the 
individual $R$ matrices corresponding to the gates.

We again consider the probabilities with which certain measurement
outcomes are obtained, i.e. $p(y^*|x)$ in Eq. (\ref{pyx1}), and show
that as before these quantities are equal to the Pfaffian of
some antisymmetric matrix.   

As before, we consider an input state $\ket{x}=a_{p_1}^{\dagger}\ldots
a_{p_l}^{\dagger} \ket{{\bf 0}}$, which we will now write as $c_{2p_1}
\ldots c_{2p_l} \ket{{\bf 0}}$ with $p_1 < p_2 < \ldots p_l$. Thus we
would like to evaluate
\beq
p(y^*|x)=\bra{{\bf 0}}c_{2p_l}\ldots c_{2p_1} \;U^{\dagger} a_{j_1} U U^{\dagger}a_{j_1}^\dagger U \ldots U^{\dagger} a_{j_k}^\dagger U U^{\dagger}a_{j_k}U\; c_{2p_1}\ldots c_{2p_l} \ket{{\bf 0}},\label{pyxbef}
\eeq
The pattern of $a$ and $a^\dagger$ is again determined by the
state $\ket{y^*}$.  We need a formula for how our general (non-number
conserving) $U$ acts by conjugation on the creation and destruction
operators.  We can use Eq. (\ref{important}):
\beq
U^{\dagger} a_iU={1\over 2}
U^{\dagger}(c_{2i}+i c_{2i+1})U=
{1\over 2}\sum_j (R_{2i,j}^T+iR_{2i+1,j}^T) c_j=
\sum_j T_{ij} c_j,
\label{defT}
\eeq
and similarly
\beq
U^{\dagger} a_i^{\dagger} U=\sum_j T_{ij}^* c_j.
\eeq
This defines the $n\times 2n$ matrix $T$.  We then obtain for the measurement
probability:
\bea
\lefteqn{p(y^*|x)=
\sum_{m_1,n_1,\ldots,m_k,n_k}T_{j_1,m_1}T_{j_1,n_1}^* \ldots T_{j_k,n_k}^{*} T_{j_k,m_k}} \nonumber \\
& & \hspace{5cm}\bra{{\bf 0}}c_{2p_l}\ldots c_{2p_1} \; c_{m_1} c_{n_1}\ldots c_{n_k}c_{m_k} c_{2p_1}\ldots c_{2p_l} \ket{{\bf 0}}.
\label{pyxgen}
\eea
Again we can use Wick's theorem to evaluate the vacuum matrix element.
This is done by writing the Majorana operators in terms of the fermion
creation and annihilation operators.  Expanding gives a large number
of terms, to each of which Wick's theorem applies.  Each term normal
orders differently, but in every case only the fully contracted terms
survive.  All of these fully contracted terms are generated by
contractions directly over the Majorana operators, defined by linear
extension:
\beq
\wick{1}{<*c_{2i}>*c_{2j+1}}=-i(
\wick{1}{<*a_i>*a_j}-
\wick{1}{<*a_i>*a_j^\dagger}+
\wick{1}{<*a_i^\dagger>*a_j}-
\wick{1}{<*a_i^\dagger>*a_j^\dagger})=
i \delta_{ij},
\eeq
And similarly
\beq
\wick{1}{<*c_{2i+1}>*c_{2j}}=-i \delta_{ij},\;\;\wick{1}{<*c_{2i}>*c_{2j}}=\wick{1}{<*c_{2i+1}>*c_{2j+1}}=\delta_{ij}.
\eeq
Here we have used Eq. (\ref{contr}).  Then, the vacuum expectation
value is written as the sum of all fully contracted expressions over
the Majorana operators, with the usual fermionic sign.  Thus, we can
say that Wick's theorem applies in the same way to the Majorana
fermion operators as it does to ordinary fermion creation and
annihilation operators; we emphasize that this is only true for the
vacuum expectation value, it is {\em not} true as an operator identity
(normal ordering is not defined for the Majorana operators).

We can summarize these contraction rules by writing $\wick{1}{<*c_i
>*c_j}=H_{ij}$ where $H$ is a $2n\times 2n$ Hermitian matrix
consisting of $2\times 2$ blocks:
\beq
H=
\left(
\ba{rrrrr} 
1 &  \ i & & &  \\
-i &  \ 1 & & & \\
& & \ddots & & \\
& & & 1 & \ i \\
& & & -i & \ 1
\ea
\right).
\eeq
Applying Wick's theorem again leads to a Pfaffian expression.  In
Table II we give entries that permit the $2(l+k) \times 2(l+k)$ matrix
$N$ to be constructed such that $p(y^*|x)={\rm Pf}(N)$.  Again, the
entire evaluation is clearly doable in polynomial time.

\section{Intermediate Measurements and Classically Conditioned Operations}
\label{ccgates}

We now extend our quantum circuit of noninteracting fermions by
allowing intermediate complete von Neumann measurements in the
computational basis on subset of qubits, which then determine the
subsequent choices of unitary gates and measurements on the remaining
qubits. We will show here that a fermionic circuit with these resources can
still be simulated efficiently with a classical algorithm. Care has to
be taken in specifying which intermediate and final measurements are
allowed in our model; we restrict ourselves to complete von
Neumann measurement in the computational basis, i.e. the outcomes of
the measurement are either `no fermion present in this mode' or `one
fermion present in this mode'.  A lot of added power can be hidden in
the kind of measurements that one is allowed to do; for example it has
been shown in Ref. \cite{kitaev:fermi} that universal quantum
computation can be achieved by noninteracting fermion gates plus a
nondestructive eigenvalue measurement of the quartic operator $c_j c_k
c_r c_s$.

\begin{figure}
\epsfxsize=16cm
\epsffile{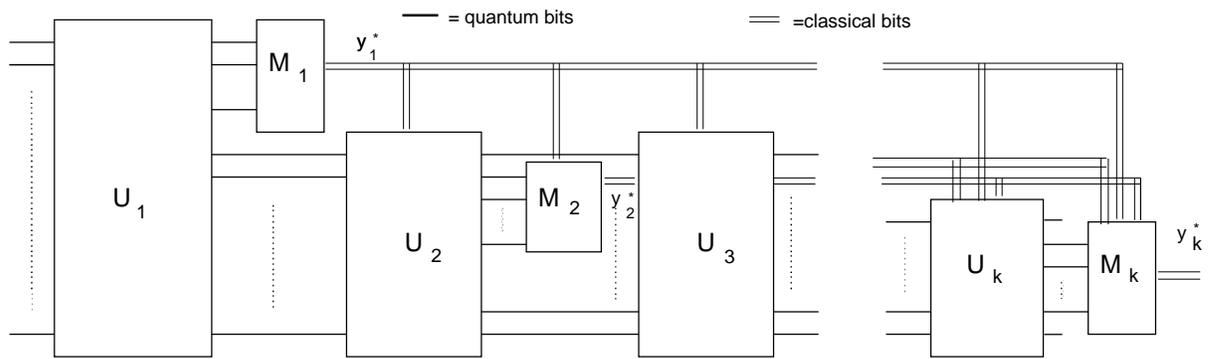}
\caption{A quantum circuit with classically conditioned gates.} 
\label{fig1}
\end{figure}

A general quantum circuit employing our set of resources is depicted
in Fig. \ref{fig1}.  Every time a measurement is made on a subset of
qubits, these qubits are no longer used in any later steps of the
computation. Our classical simulation will be constructed in the
following manner. Measurements $M_1, \ldots, M_k$ on subsets $S_1,
\ldots, S_{k}$ will take place at `times' $t_1, \ldots, t_k$.  The
total unitary evolution until the measurement $M_1$ is denoted as
$U_1$, the conditional unitary evolution between $M_k$ and $M_{k+1}$
is denoted as $U_{k+1}(y^*_1,y^*_2,\ldots,y^*_k)$ where the labels
$y^*_1,y^*_2,\ldots,y^*_k$ correspond to the outcomes of the
measurements $M_1,\ldots,M_k$.  The choice of measurements themselves
may depend on earlier measurement outcomes, i.e. $M_l=M_l(y_1^*,\ldots,y_{l-1}^*)$.
Even though the later time-evolution operators will not act on the qubits that are already measured, we keep the dimension of these matrices the same as the initial matrix $U_1$, i.e. these are $2^n \times 2^n$ matrices when the total number of fermionic modes is $n$.

We can calculate the probability that at time $t_1$ subset $S_1$ is in
the state $\ket{y^*_1}$ (and sample from this probability
distribution) by the methods that we have developed in Sections
\ref{preserve} and \ref{gensim}.  If the quantum measurement gives
$\ket{y^*_1}$, then the remaining qubits are in the state
\beq
\rho_2=\frac{U_2(y^*_1) P_{y^*_1} U_1 \ket{x} \bra{x}U_1^{\dagger} P_{y^*_1} U_2(y^*_1)^{\dagger}}{{\rm Tr} P_{y^*_1} U_1\ket{x}\bra{x}U_1^{\dagger}}, 
\label{firststep}
\eeq
where the projector $P_{y^*_1}$ is of the form
$a_{j_1}a_{j_1}^{\dagger} \ldots a_{j_{|S_1|}}^{\dagger}a_{j_{|S_1|}}$
where $j_1,\ldots,j_{|S_1|} \in S_1$, and whether the factor $a_{j_i}
a_{j_i}^{\dagger}$ or $a_{j_i}^{\dagger}a_{j_i}$ appears depends on
whether $(y^*_1)_{j_i}$ is 0 or 1 respectively.  Let us assume that we
have sampled the measurement probability distribution at time $t_1$
and have found a particular outcome $y^*_1$. To sample from the
probability distribution of measurement $M_2$, we will have to be able
to evaluate
\beq
p(y_2^*|y_1^*,x)={\rm Tr}\,P_{y^*_2} \rho_2.
\eeq
$P_{y_2^*}$, like $P_{y_1^*}$, is again a product of creation and
annihilation operators, e.g., $a_{i_1}^\dagger a_{i_1} \ldots
a_{i_{|S_2|}}^\dagger a_{i_{|S_2|}}$ where $i_1,\ldots,i_{|S_2|} \in
S_2$ and the pattern of creation and annihilation operators depends on
the bits of $y^*_2$.  The denominator in Eq. (\ref{firststep}) is
already determined when simulating the first measurement, so we will
focus on calculating
\bea
\lefteqn{p(y_1^*,y_2^*|x)=}\nonumber \\
& & {\rm Tr}\,P_{y_2^*} U_2(y^*_1) P_{y^*_1} U_1 \ket{x} \bra{x}U_1^{\dagger} P_{y^*_1} U_2(y^*_1)^{\dagger}=\bra{x} U_1^{\dagger} P_{y^*_1} U_2(y^*_1)^{\dagger}  P_{y^*_2} U_2(y^*_1) P_{y^*_1} U_1 \ket{x}.
\label{matelem}
\eea
This equation has basically the same form as Eq. (\ref{pyxbef}), except 
that (1) we have more annihilation and creation operators and (2) we conjugate 
different sets of operators by different unitary matrices. The important 
fact here is that we can again express the probability as the Pfaffian of 
some antisymmetric matrix. Let us see how we construct this matrix.

At this point we simplify the notation somewhat. Let $U_2(y^*_1)=U_2$,
$P_{y^*_1}=P_1$ and $P_{y^*_2}=P_2$. We put in $U_1 U_1^{\dagger}$ and $U_2
U_2^{\dagger}$ terms in the appropriate places in Eq. (\ref{matelem}),
so that operators in the first (from the left) $P_1$ get conjugated
by $U_1^{\dagger}$, operators in $P_2$ get conjugated by
$U_{12}^{\dagger} \equiv U_1^{\dagger}U_2^{\dagger}$ and the last $P_1$ gets
conjugated by $U_1^{\dagger}$ again. Let $T^k$, $k=1,12$, be defined by 
$U_k^{\dagger} a_i U_k=\sum_j T_{ij}^k c_j$. We obtain:
\bea
\lefteqn{p(y_1^*,y_2^*|x)=
\sum_{\stackrel{a_1,b_1,f_1,g_1,\ldots,a_{|S_1|},b_{|S_1|},f_{|S_1|},g_{|S_1|}}{d_1,e_1,\ldots,d_{|S_2|},e_{|S_2|}}} 
T_{j_1,a_1}^1 T_{j_1,b_1}^{1\,*} \ldots 
T_{j_{|S_1|},b_{|S_1|}}^{1\,*} T_{j_{|S_1|},a_{|S_1|}}^1 
} \nonumber \\
& & 
T_{i_1,e_1}^{12\,*} T_{i_1,d_1}^{12}  \ldots 
T_{i_{|S_2|},e_{|S_2|}}^{12\,*} T_{i_{|S_2|},d_{|S_2|}}^{12}
T_{j_1,f_1}^1 T_{j_1,g_1}^{1\,*} \ldots 
T_{j_{|S_1|},g_{|S_1|}}^{1\,*} T_{j_{|S_1|},f_{|S_1|}}^1 \\
& & \bra{{\bf 0}}c_{2p_l}\ldots c_{2p_1} \; (c_{a_1} c_{b_1}\ldots c_{b_{|S_1|}}c_{a_{|S_1|}})(c_{e_1} c_{d_1}\ldots c_{e_{|S_2|}}c_{d_{|S_2|}}) 
(c_{f_1} c_{g_1}\ldots c_{g_{|S_1|}}c_{f_{|S_1|}})c_{2p_1}\ldots c_{2p_l} \ket{{\bf 0}}.\nonumber
\label{condpyxgen}
\eea
In Table III we show how to construct the matrix $O$ of dimension
$2(l+2|S_1|+|S_2|)$ for which $p(y_1^*,y_2^*|x)={\rm Pf}(O)$. The
notation $c_{(a/f)_{\alpha}}$ indicates that the $c$-operator can be
either a $c_{a_{\alpha}}$ or a $c_{f_{\alpha}}$; the reason is that
these operators have identical $T$ prefactors.

It is clear that we can extend this procedure to the case of a circuit
that contains $k={\rm poly}(n)$ instances of measurements on subsets
that determine the next choice of unitary evolution. In general when
we express a probability such as Eq. (\ref{matelem}), we see that
$P_1$ gets conjugated by $U_1$, $P_2$ by $U_{12}$, $P_3$ by
$U_{123}$ ,..., and $P_k$ by $U_{12\ldots k}$, the total unitary
evolution. When we write $p(y_1^*, y_2^* \ldots y_k^*|x)={\rm Pf}(X)$,
the dimension of the matrix $X$ is
$2(l+|S_k|+2\sum_{i=1}^{k-1}|S_{i}|)$.  The entries of this matrix can
be determined by calculating particular matrix elements (specified by
the measured sets of qubits) of at most $(2k+1)^2$ matrices of the form $T^i H
{T^j}^{\dagger}$ etc. where $i$ and $j$ are labels which can be
$1,12,123,\ldots, 123\ldots k$.

Let us summarize the simulation algorithm:
\vspace{1cm}

{\bf Classical simulation of a quantum circuit with noninteracting fermions 
and fermion counting measurements, see Fig. \ref{fig1}}:\\
\begin{enumerate}
\item Compute the $n\times 2n$ matrix $T^1$ corresponding to
$U_1^{\dagger}$, Eq. (\ref{defT}).
\item Simulate measurement $M_1$: sample from the probability
distribution $p(y_1^*|x)$ using the measurement theorem in
Ref. \cite{valiant:simulation} and the fact that $p(y^*|x)={\rm
Pf}(O_1)=\sqrt{\det O_1}$ where $O_1$ is a $2(l+k) \times 2(l+k)$
matrix with $k$ equal to the Hamming weight of input string $x$ and
$l$ equal to the number of bits in $y_1^*$.
\item Let $y_1^*$ be the outcome of this measurement $M_1$, and
let $U_2$ be the corresponding unitary evolution.  Compute $T^{12}$
corresponding to $U_1^{\dagger}U_2^{\dagger}$.
\item Simulate measurement $M_2$: sample from the probability
distribution $p(y_2^*|y_1^*,x)= \frac{p(y_1^*,y_2^*|x)}{p(y_1^*|x)}$
where we use the fact that we can evaluate $p(y_1^*,y_2^*|x)={\rm
Pf}(O_2)$.  $O_2$ depends on $T^{12}$ and $T^1$ as in Table III.
\item Let $y_2^*$ be the outcome of the measurement $M_2$ and let
$U_3$ be the corresponding unitary evolution, possibly also depending
on the first outcome $y_1^*$.  Calculate $T^{123}$ corresponding to
$U_1^{\dagger}U_2^{\dagger}U_3^{\dagger}$.
\item Simulate measurement $M_3$: sample from the probability
distribution
$p(y_3^*|y_1^*,y_2^*,x)=\frac{p(y_1^*,y_2^*,y_3^*|x)}{p(y_1^*,y_2^*|x)}$,
where we use the fact that we can evaluate $p(y_1^*,y_2^*,y_3^*|x)={\rm
Pf}(O_3)$. $O_3$ depends on $T^{123}$, $T^{12}$ and $T^1$.
\item Repeat steps 5 and 6 for the subsequent evolutions
$U_4,\ldots,U_k$, finding expressions for $T^{123 \ldots k}$, and
finally simulate the last measurement $M_k$ by sampling from the
distribution $p(y_k^*|y_1^*,y_2^*,\ldots, y_{k-1}^*,
x)=\frac{p(y_1^*,y_2^*,\ldots, y_k^*|x)}{p(y_1^*,y_2^*,\ldots
y_{k-1}^*|x)}$.
\end{enumerate}

 
It is evident that this procedure is polynomial when the number of stages
$k$ of the compute/measure procedure of Fig. 1 is poly($n$): the largest
matrix whose Pfaffian must be computed has dimension bounded by $4kn$.

\section{Discussion}
\label{disc}

The present work opens a set of very interesting questions concerning
the boundary between classical and quantum computation.  For fermionic
quantum circuits we may ask, what is the effect of adding circuit
elements beyond those considered above (those associated with a
noninteracting fermion model)?  Three outcomes are possible: 1) the
circuit can perform universal quantum computation, 2) the circuit
remains efficiently simulatable by a classical computation, or 3) some
intermediate case. For example, one could explore the effect of adding 
(unphysical) linear terms to the gate Hamiltonians. These 
terms $a_i+a_i^{\dagger}$ and $i(a_i-a_i^{\dagger})$ will be somewhat 
similar, but not identical to 1-qubit gates in terms of Pauli-matrices; 
they are nonlocal gates as can be seen from the Jordan-Wigner transformation, Eq. (\ref{jw}). It can in fact be shown \cite{knill:flo} that these 
linear interactions can be incorporated in purely quadratic fermion interactions by adding a new fermionic mode, which we may label $''-1''$, and changing the linear interactions on, say,
mode $i$ to quadratic interactions between mode $i$ and mode $-1$. Another line of investigation could be into some known physical models (e.g., the Anderson model and the Kondo model \cite{Ashmer}) that involve more general fermionic interactions at a single site (meaning $2$ fermionic modes), many of whose properties are computable. 


Valiant's work shows that some terms
added to the fermion model (or some gates added to the circuit model)
still result in a classically simulatable system; these new terms are
to act on the first two fermionic modes only. These new gates (if they 
exist) do not preserve the parity of the number of fermions, and thus 
involve either linear or cubic terms in the annihilation and creation 
operators. (This follows from the fact that any gate that obeys the 
5 matchgate identities in Ref. \cite{valiant:simulation} and preserves 
the parity of the number of fermions, is automatically quadratic in 
the number of fermions.) See Ref. \cite{knill:flo} for a more extensive
treatment of these additional gates.


The case of adding power by intermediate measurements is also quite
interesting: between the case of complete von Neumann measurements 
that are classically simulatable, and the quartic-operator basis measurements of Bravyi and Kitaev \cite{kitaev:fermi}, which give universal quantum computation, there are many possible POVM measurement scenarios that have not been
analyzed. We are hopeful that further analysis will be able to identify more scenarios as definitely classical or definitely quantum.

This classical-quantum boundary is remarkably different for fermionic
and bosonic systems.  A model quadratic in bosonic operators describes
the ``linear optics'' scenario of quantum computation; formally, this model
only differs from the noninteracting-fermion model in that bosons are
characterized by commutation rather than anticommutation relations:
\bea
[a_i,a_j]\equiv a_i a_j-a_j a_i=0,\;\;&\;\; [a_i^{\dagger},a_j^{\dagger}]=0,\;\; &\;\; [a_i,a_j^{\dagger}]=\delta_{ij}I.
\label{crel}
\eea
Nevertheless, the exact parallel to the model we analyzed above, in
which quadratic Hamiltonians can be interspersed with complete von
Neumann measurements, is fully ``quantum'' in the bosonic case
\cite{klm} despite being ``classical'' for fermions.  


We would like to emphasize again that this quantum/classical
distinction may not be perfectly sharp; being able to efficiently
compute some properties of a circuit classically does not mean that
every aspect of the quantum dynamics of this circuit are also
efficiently computable.  This is shown when we try to carry out the
analysis in Section \ref{preserve} and Section \ref{gensim} for
bosons, to see where the parallels between the two cases break down.
In fact, one can go surprisingly far before any differences appear.
An equation of the form of Eq. (\ref{conju}) still applies for bosons,
where $V$ can be be shown to be any element in $U(n)$ \cite{reck}.
That is, a canonical transformation exists in both the boson and
fermion case which permits a representation of the system as a set of
noninteracting particles.  

It might have been thought that this easy diagonalization leads to an
easy computation of all possible properties, but this does not appear
to be the case.  An essential difference is that no sign changes occur
when we interchange the bosonic creation operators amongst each other.
This causes an expression such as $\bra{y}U \ket{x}$ in Section
\ref{preserve} to be equivalent to the {\em permanent} of some matrix,
if we analyze the case when $\ket{x}$ and $\ket{y}$ both contain no
more than one boson per mode. The permanent is a much harder
object to calculate exactly than the determinant of the fermion case,
in fact this has been proved to be an $\#P$-complete problem
\cite{valperm}. Our methods will therefore fail to evaluate these
bosonic matrix elements efficiently.

In summary, our results on the classical simulatability of
noninteracting fermions leave some interesting questions unanswered
about the computational power of various physical models of quantum
computation.  While Valiant's analysis turns out to conform
largely to a ``known'' area of physics, his work shows that
mathematical approaches to these problems are possible that have never
been envisioned in many-body physics.  


\section*{Acknowledgments}

We are grateful for the support of the National Security Agency and the
Advanced Research and Development Activity through Army Research
Office contract numbers DAAG55-98-C-0041 and DAAD19-01-C-0056.  We
thank M. Devoret, E. Knill, and R. Laflamme for helpful discussions.
We also thank E. Knill for showing us his unpublished manuscript on 
fermions and matchgates.


\begin{table}[h]
\begin{tabular}[h]{l|l||c|c|c|c}
\multicolumn{1}{l}{ } & & \multicolumn{4}{c}{{\em j}}\\
\cline{3-6}
\multicolumn{1}{l}{ } & & $a_{p_{\beta}}$ & $a_{n_{\beta}}$ & $a_{m_{\beta}}^{\dagger}$ & $a_{p_{\beta}}^{\dagger}$\\
\hline
\hline
& $\,a_{p_{\alpha}}$ & 0 & 0 & $V^{\dagger}_{j_{\beta},p_{\alpha}}$ & $\delta_{\alpha,\beta}$ \\
\cline{2-6}
{\em i}\ & $\,a_{n_{\alpha}}$ & X & $\sum_{n_{\alpha},n_{\beta}} V_{n_{\alpha},j_{\alpha}} V_{n_{\beta},j_{\beta}}\wick{1}{<*a_{n_{\alpha}}>*a_{n_{\beta}}}=0$ & $\sum_{n_{\alpha},m_{\beta}} V_{n_{\alpha},j_{\alpha}} V_{j_{\beta},m_{\beta}}^{\dagger}\wick{1}{<*a_{n_{\alpha}}>*a_{m_{\beta}}^{\dagger}}=\delta_{\alpha,\beta}$ & $V_{p_{\beta},j_{\alpha}}$ \\
\cline{2-6}
& $\,a_{m_{\alpha}}^{\dagger}$ & X & $\sum_{m_{\alpha},n_{\beta}} V^{\dagger}_{j_{\alpha},m_{\alpha}} V_{n_{\beta},j_{\beta}} \wick{1}{<*a_{m_{\alpha}}^{\dagger} >*a_{n_{\beta}}}=0$ & $\sum_{m_{\alpha},m_{\beta}}V^{\dagger}_{j_{\alpha},m_{\alpha}}V^{\dagger}_{j_{\beta},m_{\beta}}\wick{1}{<*a^{\dagger}_{m_{\alpha}} >*a^{\dagger}_{m_{\beta}}}=0$ & 0 \\
\cline{2-6}
& $\,a_{p_{\alpha}}^{\dagger}$ & X & X & X & 0 \\
\cline{2-6}
\end{tabular}
\vspace{0.5cm}
\caption{The matrix $M(i,j)$ for $i < j$.}
\end{table}

\begin{table}[h]
\begin{tabular}[h]{l|l||c|c|c}
\multicolumn{1}{l}{ } & & \multicolumn{3}{c}{{\em j}}\\
\cline{3-5}
\multicolumn{1}{l}{ } & & $c_{m_{\beta}}$ & $c_{n_{\beta}}$ & $c_{2p_{\beta}}$ \\
\hline
\hline
& $c_{m_{\alpha}}$ & $(T H T^T)_{j_{\alpha},j_{\beta}}$ & $(T H T^{\dagger})_{j_{\alpha},j_{\beta}}$  & $(TH)_{j_{\alpha},2p_{\beta}}$\\
\cline{2-5}
{\em i}\ & $c_{n_{\alpha}}$ & $(T^*HT^T)_{j_{\alpha},j_{\beta}}$ & $(T^*HT^{\dagger})_{j_{\alpha},j_{\beta}}$ & $(T^*H)_{j_{\alpha},2p_{\beta}}$ \\
\cline{2-5}
& $c_{2p_{\alpha}}$ & $(HT^T)_{2p_{\alpha},j_{\beta}}$ & $(HT^{\dagger})_{2 p_{\alpha},j_{\beta}}$ & $\delta_{\alpha,\beta}$ \\
\cline{2-5}
\end{tabular}
\vspace{0.5cm}
\caption{The matrix $N(i,j)$ for $i < j$.}
\end{table}

\begin{table}[h]
\begin{tabular}[h]{l|l||c|c|c|c|c}
\multicolumn{1}{l}{ } & & \multicolumn{5}{c}{{\em j}}\\
\cline{3-7}
\multicolumn{1}{l}{ } & & $c_{(a/f)_{\beta}}$ & $c_{(b/g)_{\beta}}$ & $c_{d_{\beta}}$ & $c_{e_{\beta}}$ & $c_{2p_{\beta}}$ \\
\hline
\hline
& $c_{(a/f)_{\alpha}}$ & $(T^1 H {T^1}^T)_{j_{\alpha},j_{\beta}}$ & $(T^1 H {T^1}^{\dagger})_{j_{\alpha},j_{\beta}}$ & $(T^1 H {T^{12}}^T)_{j_{\alpha},i_{\beta}}$ & $(T^1 H {T^{12}}^{\dagger})_{j_{\alpha},i_{\beta}}$ & $(T^1H)_{j_{\alpha},2p_{\beta}}$\\
\cline{2-7}
& $c_{(b/g)_{\alpha}}$ & $({T^1}^* H {T^1}^T)_{j_{\alpha},j_{\beta}}$ & $({T^1}^*H{T^1}^{\dagger})_{j_{\alpha},j_{\beta}}$ & $({T^1}^* H {T^{12}}^T)_{j_{\alpha},i_{\beta}}$   &$({T^1}^*H{T^{12}}^{\dagger})_{j_{\alpha},i_{\beta}}$  & $({T^1}^* H)_{j_{\alpha},2p_{\beta}}$ \\
\cline{2-7}
{\em i}\ & $c_{d_{\alpha}}$ & $(T^{12} H {T^1}^T)_{i_{\alpha},j_{\beta}}$ & $(T^{12} H {T^1}^{\dagger})_{i_{\alpha},j_{\beta}}$ & $(T^{12} H {T^{12}}^T)_{i_{\alpha},i_{\beta}}$    & $(T^{12} H {T^{12}}^{\dagger})_{i_{\alpha},i_{\beta}}$  & $(T^{12}H)_{i_{\alpha},2p_{\beta}}$\\
\cline{2-7}
& $c_{e_{\alpha}}$ & $({T^{12}}^* H {T^1}^T)_{i_{\alpha},j_{\beta}}$ & $({T^{12}}^* H {T^1}^{\dagger})_{i_{\alpha},j_{\beta}}$&  $({T^{12}}^* H {T^{12}}^T)_{i_{\alpha},i_{\beta}}$  & $({T^{12}}^* H {T^{12}}^{\dagger})_{i_{\alpha},i_{\beta}}$   &$({T^{12}}^*H)_{i_{\alpha},2p_{\beta}}$ \\
\cline{2-7}
& $c_{2p_{\alpha}}$ & $(H{T^1}^T)_{2p_{\alpha},j_{\beta}}$ & $(H{T^1}^{\dagger})_{2 p_{\alpha},j_{\beta}}$ & $(H{T^{12}}^T)_{2p_{\alpha},i_{\beta}}$  & $(H{T^{12}}^{\dagger})_{2 p_{\alpha},i_{\beta}}$  & $\delta_{\alpha,\beta}$ \\
\cline{2-7}
\end{tabular}
\vspace{0.5cm}
\caption{The matrix $O(i,j)$ for $i < j$.}
\end{table}

\bibliographystyle{hunsrt}
\bibliography{refs}

\end{document}